# The Wide-Field Imaging Interferometry Testbed I: purpose, testbed design, data, and synthesis algorithms

David Leisawitz[*a], Bradley J. Frey[a], Douglas B. Leviton[a], Anthony J. Martino[a], William L. Maynard[a], Lee G. Mundy[b], Stephen A. Rinehart[a,c], Stacy H. Teng[b], Xiaolei Zhang[d]

[a]NASA Goddard Space Flight Center, [b]Department of Astronomy, University of Maryland, [c]National Research Council Research Associate, [d]SSAI

**ABSTRACT**

The Wide-field Imaging Interferometry Testbed (WIIT) was designed to validate, experiment with, and refine the technique of wide field mosaic imaging for optical/IR interferometers. Here we offer motivation for WIIT, present the testbed design, and describe algorithms that can be used to reduce the data from a spatial and spectral Michelson interferometer. A conventional single-detector Michelson interferometer operating with narrow bandwidth at center wavelength $\lambda_c$ is limited in its field of view to the primary beam of the individual telescope apertures, or $\sim \lambda_c / d_{tel}$ radians, where $d_{tel}$ is the telescope diameter. Such a field is too small for many applications; often one wishes to image extended sources. We are developing and testing techniques analogous to the mosaicing method employed in millimeter and radio astronomy, but applicable to optical/IR Michelson interferometers, in which beam combination is done in the pupil plane. An $N_{pix} \times N_{pix}$ array detector placed in the image plane of the interferometer is used to record simultaneously the fringe patterns from many contiguous telescope fields, effectively multiplying the field size by $N_{pix}/2$, where the factor 2 allows for Nyquist sampling. This technique will be especially valuable for interferometric space observatories, such as the Space Infrared Interferometric Telescope (SPIRIT) and the Submillimeter Probe of the Evolution of Cosmic Structure (SPECS). SPIRIT and SPECS will be designed to provide sensitive, high angular resolution, far-IR/submillimeter observations of fields several arcminutes in diameter, and views of the universe complementary to those provided by the Hubble Space Telescope (HST), the Next-Generation Space Telescope (NGST), and the Atacama Large Millimeter Array (ALMA).

## 1. INTRODUCTION

In most modern optical/IR interferometers[1] beam combination is done in the pupil plane instead of the image plane. Pupil plane beam combination also appears to be the more attractive alternative for space-based IR wavelength interferometry, for reasons that will be given below. This approach, which is conventionally called the "Michelson" method, has several advantages but one often-cited disadvantage: a classical Michelson interferometer has a narrow field of view. Can this drawback be overcome? Can an analog to the mosaicing techniques employed at radio and millimeter wavelengths[2] be developed for applications at shorter wavelengths, where the most sensitive detection methods preserve no phase information and the beams must be directly combined? If so, then the scientific productivity of future space interferometers can be significantly enhanced.

The Wide-Field Imaging Interferometry Testbed (WIIT) was assembled at NASA's Goddard Space Flight Center to develop the technique of wide-field mosaic imaging for optical/IR Michelson interferometers. Beginning with fundamental astrophysics questions we summarize the derived measurement requirements for the far-IR/sub-mm spectral region and give the motivation for this project in the next section. The specific goals of the WIIT project are described in Section III, and our approach to the problem is given in Section IV. We describe the testbed design in Section V and give an overview of the WIIT data products in Section VI.

---

Two companion papers in this proceedings volume further describe WIIT. Paper II[3] describes the testbed in more detail, presents measurements that characterize the performance of the instrument, and outlines future plans. Paper III[4] describes the WIIT metrology system, which has several novel aspects. The current series of papers update our progress since early descriptions of the testbed design and research plan.[5, 6]

## 2. MOTIVATION

Some of the most compelling unanswered questions in cosmology and astrophysics will lie dormant or, at best, be unsatisfactorily addressed until measurement capabilities in the far-infrared and submillimeter spectral region are vastly improved. For example, we would like to measure the cosmic history of star and galaxy formation and the buildup of heavy elements over time, but dust extinction hides many of the interesting sources or plagues UV/optical measurements with large uncertainty. We would like to study the astrophysical processes associated with star and planet formation, but these processes also occur behind veils of dust. Far-IR/sub-mm observations are largely immune to extinction, and the sources are intrinsically bright and interesting at these wavelengths. Existing observations from the Infrared Astronomical Satellite (IRAS), the Cosmic Background Explorer (COBE), and the Infrared Space Observatory (ISO) have whetted the scientific community's appetite for better far-IR telescopes. Though very powerful, the next-generation observatories Space IR Telescope Facility (SIRTF), Herschel, and the Stratospheric Observatory for Infrared Astronomy (SOFIA) will still fall short of the sensitivity and resolution capabilities needed to exploit fully this information-rich spectral region.

The desired measurement capabilities are summarized in Table 1.[7] The sensitivity requirement is dictated by the need to detect sources at cosmological distances. Further, we will want to: (a) observe them with spectral resolution at least high enough to detect line emission ($R = \lambda/\Delta\lambda > 1000$ for an unresolved galaxy) or, ideally, measure internal Doppler motions ($R > 10^5$ for protostars and protoplanetary systems); (b) image with high angular resolution, initially to break the extragalactic source confusion barrier (~1 arcsec), and eventually to resolve distant galaxies and nearby protoplanetary systems at a level comparable to that of HST or NGST: tens of milli-arcseconds; and (c) observe a large number of objects in the lifetime of a space mission.

Table 1. Desired Measurement Capabilities for the Mid-IR to Millimeter Spectral Range

| Science goal | Formation and evolution of cosmic structure | Formation of stars and planetary systems |
|---|---|---|
| Sample targets | Hubble Deep Fields, gravitational lens sources, interacting galaxies | Nearest protostars, Orion prolyds, Vega, HH 30, and other disks |
| Wavelength range (peak emission) (μm) | 40 – 1000 | 30 – 300 |
| Angular resolution (mas) | 20 | 10 |
| Spectral resolution ($\lambda/\Delta\lambda$) | $>10^4$ | $3 \times 10^5$ |
| Point source sensitivity, $\nu S_\nu$ (W/m$^2$) | $10^{-20}$ | $10^{-20}$ |
| Field of view (arcmin) | 4 | 4 |

Ongoing detector, cryocooler, and large-mirror technology research will enable us to build background-limited far-IR telescopes capable of detecting galaxies back to the epoch of their formation.[8, 9] A background-limited telescope with total aperture area in the tens of square meters would provide ample sensitivity in a reasonable exposure time.

If the collecting area were deployed as a single-aperture telescope, the first angular resolution goal (to beat confusion) would be satisfied at wavelengths shorter than about 100 μm. To obtain the same resolution at longer wavelengths, or to achieve the more demanding resolution goal (matching HST and NGST), the light collecting area can be partitioned among subapertures, which can be used as elements of an interferometer. If the sub-apertures are separated by a distance

(baseline) whose maximum length is $b_{max}$, the resolution of the interferometer will be $\theta_{res} = 10$ mas $(\lambda/100$ μm$)(b_{max}/1$ km$)^{-1}$. A 1 km maximum baseline is needed to provide Hubble-class resolution in the far-IR/sub-mm.

Recognizing the science potential and the technical possibilities summarized above, the astrophysics community's desire for a new generation of far-infrared space telescopes after SIRTF and Herschel, beginning with the Single-Aperture Far-IR (SAFIR) telescope,[10] and followed by a far-IR/submillimeter interferometer, was stated in the "Decadal Report".[11] In addition to SAFIR, the SPIRIT and SPECS far-IR/sub-mm interferometers now under study by NASA were conceived to accomplish the scientific objectives mentioned above,[12, 13] SPIRIT being a science and technology pathfinder for the more ambitious SPECS mission.

For several reasons SPIRIT and SPECS will likely use Michelson rather than Fizeau beam combination. First and foremost, a Fizeau interferometer would need a large-format detector array to sample the fringe pattern in the spatial domain, whereas a Michelson interferometer, which measures fringes in the time domain, needs very few detectors. Given the current state of far-IR detector and array readout device development it would be reasonable to assume that a $100 \times 100$ pixel detector array could be available in a decade, but not a much larger array; the market for such detectors is very limited. Second, a Michelson interferometer can easily be operated in "double Fourier" mode,[14] so it would naturally provide the desired spectral as well as spatial resolution. By stroking the Michelson delay line to produce an optical path difference $\Delta$ one can obtain a spectrum with resolution $R = 10^4 (\Delta/1$ m$) (\lambda/100$ μm$)^{-1}$ at wavelength $\lambda$. The required physical stroke length is $\Delta/2$ if a single retro-reflector is used, and can be made smaller by increasing the number of reflections. Third, a Michelson interferometer has two complementary output ports, one of which (or a short-wavelength channel at one port) could be used for phase tracking.

An interferometer with n sub-apertures has n(n-1)/2 baselines and twice this number of output ports if it is a Michelson type.[15] This has the potential to make a Michelson beam combiner rather complex. However, SPIRIT and SPECS would use only two or three light collecting telescopes because each one would have to be cryogenic, and would therefore be expensive. The number of instantaneous baselines would be 1 or 3, depending on whether n = 2 or 3, and the beam combining optics would not be very complicated. To obtain good image quality all spatial frequencies would have to be sampled in two dimensions; in other words, measurements would have to be made on many baselines $b < b_{max}$, and at many baseline position angles. This so-called "u-v plane"[†] filling is accomplished with ground-based interferometers by deploying many apertures, allowing for array reconfiguration, and relying on Earth rotation. On the ground, at least three telescopes are needed to measure the "closure phase," a quantity free of the adverse effects of air turbulence. In space, there is no atmosphere to cause wavefront distortion, and more freedom to move apertures to desired locations, so imaging can be accomplished with a an array of very few elements, and one can tailor the u-v coverage to the problem at hand.

To achieve their primary science goals, the SPIRIT and SPECS interferometers will have to image fields comparable in size to the Hubble Deep Field, a 2.5 arcmin image captured with the Hubble Wide-Field/Planetary Camera. A conventional Michelson interferometer constructed from sub-aperture telescopes of diameter $d_{tel}$ images a region corresponding to the main diffraction lobe (Airy disk) of those telescopes; the field of view $\theta_{fov} \sim \lambda/d_{tel}$. As illustrated in Figure 1, a hypothetical far-IR/sub-mm interferometer constructed from 4 m telescopes and operating at 200 μm would have a field of view of ~0.2 arcmin.

Given that this is much smaller than the desired field size, how would we address requirement (c) to observe a large number of objects? This challenging question, coupled with the possibility that a detector array with a relatively modest number of pixels might be able to be used in a Michelson interferometer to provide spatial multiplexing, became the motivation for our development of the Wide-field Imaging Interferometry Testbed (WIIT).

---

[†] u and v are orthogonal spatial frequency components, which are measured in cycles per radian and related by $b_x/\lambda$ and $b_y/\lambda$, respectively, to orthogonal components of the baseline vector whose magnitude is $b^2 = b_x^2 + b_y^2$, and to the observed wavelength $\lambda$. The x-y plane is oriented normal to the target of observation, which in the wide-field case is the center of the field. The SPIRIT and SPECS collector mirrors move in the x-y plane. See Thompson, Moran, and Swenson (1986) for a full description of this reference frame.

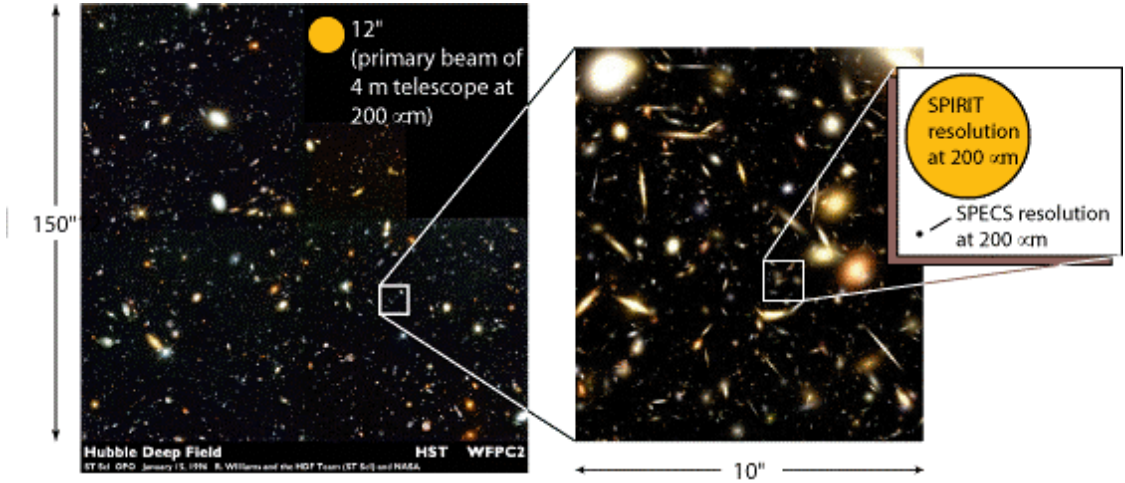

**Figure 1** – An interferometer constructed from 4 m diameter sub-apertures operating at a wavelength of 200 μm would have a 12 arcsec field of view, much smaller than the fields of astrophysical interest, such as the Hubble Deep Field (left). The image on the right is a simulation of what NGST might see in the near-IR (credit: A. Benson) in a 10 arcsec region. The rightmost inset corresponds to a 1 arcsec region, the level of resolution needed to distinguish the emissions of individual distant galaxies and protogalactic objects. SPIRIT would provide sub-arcsecond resolution in the far-IR and submillimeter, and SPECS would provide HST-class resolution in the tens of milli-arcseconds. The technique of wide-field imaging interferometry is needed to enable SPIRIT and SPECS to image several arcminute fields.

## 3. GOALS OF THE WIIT PROJECT

Our chief aspiration is to demonstrate that an optical/IR Michelson interferometer equipped with a multi-pixel detector array can image a complex, extended scene over a wide field of view (i.e., $\theta_{fov} \gg \lambda/d_{tel}$). There is no reason in principle that this cannot be done, but it has not been done before and there are many practical challenges.

Three high-level project goals derive from the desire to develop a technique for wide-field imaging suited to space-based interferometry:
1. we must develop new algorithms applicable to optical/IR "double Fourier" interferometry;
2. we must obtain representative data in order to test the algorithms; and
3. we need a Michelson interferometer with an appropriately controlled optical delay line to obtain the data.

The mosaicing algorithms developed for applications where coherent detection is possible, at millimeter and radio wavelengths, are not directly applicable to optical/IR interferometry. However, we do wish to take advantage of the thoroughly developed field of radio image synthesis[2] and preserve the mosaicing concept of "joint deconvolution".[16, 17] The new algorithms[18] should be compatible with data gathered "on the fly," as the array elements and optical delay lines of space-based imaging interferometers will almost always be moving. Further, they should derive the phase information needed to reconstruct an image from celestial sources in the science field of view and metrology internal to the instrument. This leads to an ancillary question: What are the characteristics (spatial distribution, brightness, astrometric and brightness stability, wavelength of peak emission) of a good phase reference source?

Alternative algorithms must be evaluated for their performance, both in terms of computational speed and quality of image synthesis results. What works best on irregularly sampled data and seems compatible with predicted growth in computational capability?

Real data are needed to test the algorithms. Test scenes ranging in complexity from simple pinholes to panchromatic images with complex spatial structure should be observed.

## 4. THE APPROACH

Many optical Michelson interferometers exist or are under development,[1] but none provides full u-v plane coverage and has a Fourier Transform Spectroscopy (FTS) mode for double Fourier interferometry. To obtain the data needed to accomplish the objectives described above, we could have built either an instrument for an existing telescope or a laboratory testbed interferometer. We opted for the latter because it is less costly, provides greater flexibility and accessibility, and has no significant disadvantages. By placing a source at the focus of a collimating mirror we generate a planar wavefront, the equivalent of a source in the sky. We decided to work at optical rather than IR wavelengths because the critical components (e.g., detectors) are cheaper and easier to handle, and the interferometer is smaller for the same number of resolution elements.

### 4.1. The Testbed

The testbed interferometer, WIIT, is located at NASA's Goddard Space Flight Center in a laboratory adjacent to the Diffraction Grating Evaluation Facility (DGEF). Design and assembly of the testbed took place over a two-year period beginning in May 1999, and culminated in detection of the first white light fringes (Fig. 2) about a year ago. The design is shown schematically in Figure 3, and a matching photograph of WIIT is shown in Figure 4. A light source on the right illuminates a 21-inch diameter parabolic mirror (2.4 m focal length; left, not shown) through a pinhole or a miniature scene (Fig. 5) located at the focus.[‡] A collimated beam returns to the collector mirrors (left), of which there are two. These mirrors are mounted on a stable rail and move on a cushion of air, remaining very well aligned (see Paper II), to provide baselines ranging in length from 1 mirror diameter (25 mm) to about 200 mm. The collector mirrors are elliptical, but appear circular in projection when viewed from the collimating mirror. The baseline rail is aligned perpendicular to the incident beam.[§]

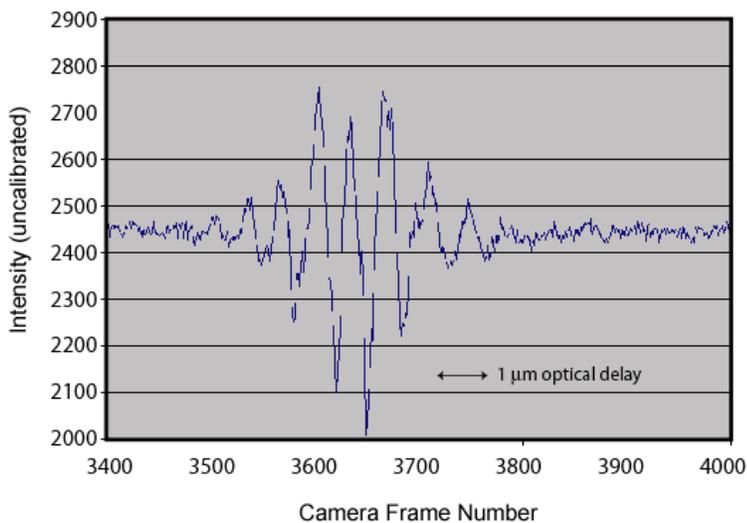

**Figure 2** – The first white light fringes were detected with WIIT on 16 August 2001. Efforts since that time (see Paper II) have led to significant improvements in data quality.

The collector mirrors reflect two pieces of the wavefront in opposite directions, combining them in a half-reflective/half-transmissive beamsplitter. Before combination, one beam passes through an optical delay line. The delay line is constructed from a pair of mirrors arranged in a rooftop configuration mounted on a commercial air-bearing translation stage. Flat mirrors are used throughout the interferometer, from the collectors to the beamsplitter. The interferometer mirrors are aligned such that the collimated beams from a source located on the optical axis of the parabolic mirror arrive at the beamsplitter exactly 90° apart, and at 45° angles to the half-reflective surface. After combination, these rays enter a lens parallel to its optical axis and are focused onto a CCD camera with a 9 μm pixel pitch. A scene placed at the focus of the collimating mirror appears as an image on the camera when either arm of the interferometer is blocked, and the resolution of the image is that given by the diffraction limit of the 25 mm diameter mirrors used in the interferometer.

---

[‡] These scenes are recorded microlithographically on a custom-maded glass plate.
[§] This configuration should be reminiscent of Michelson's *stellar* interferometer (Michelson 1920), but the analogy ends there; the collector mirrors in WIIT steer the beams apart, combining them downstream in the pupil plane (the classical "Michelson interferometer"), whereas Michelson directed starlight into the 100" telescope on Mount Wilson and used Fizeau beam combination in his stellar interferometer, unintentionally confusing subsequent generations of newcomers to the field of optical interferometry.

Various features now present in WIIT, most notably the optical encoders and rotation stages described in Paper III, have been added since the photograph in Figure 4 was taken. Two parallel filter wheels were added between the beamsplitter and the lens, allowing selection from a set of interference filters to vary the bandwidth, and from a set of neutral density filters to vary the transmitted fraction.

To understand the operation of the testbed, first consider the signal recorded as a function of time by a single camera pixel, specifically a pixel $(x_o, y_o)$ that senses light rays incident on the lens parallel to its optical axis. For a given collector mirror position (baseline), the delay line is scanned through the zero path difference (ZPD) point along a distance covering the entire white light fringe pattern. Meanwhile, the camera, responding to a trigger pulse generated by a Zygo interferometer (bottom of Figures 3 and 4), is read out at fixed optical delay intervals. The data read from pixel $(x_o, y_o)$ during the scan constitute an interferogram whose Fourier transform is a spectrum, namely the spectrum of the source at the spatial frequency sampled by the baseline. The fringe spacing in the interferogram is comparable to the wavelength at the center of the bandpass, $\lambda_c \sim 600$ nm. The camera is triggered at a selectable optical delay interval, and the interval is typically set in the tens of nm range, yielding a densely sampled fringe pattern. Next the collecting mirrors are moved to change the baseline length and a new scan is taken. A series of such scans are made with baseline lengths set to densely sample the source at all spatial frequencies up to a maximum frequency corresponding to the maximum baseline length.

The resulting single-pixel data set contains all the information needed to reconstruct a 1-dimensional image of the source. Two-dimensional image synthesis is accomplished by rotating the source to gain access to baselines having a vector component perpendicular to the collector mirror rail. For each rotation angle the baseline and delay line mirrors are moved as described above. The reconstructed image has a resolution given by the "synthesized beam" diameter

$$\theta_s = \lambda_c / 2b_{max} \approx 600 \text{ nm} / 400 \text{ mm} = 1.5 \times 10^{-6} \text{ rad} \approx 0.3 \text{ arcsec} \qquad (1)$$

and covers a field of view given by the "primary beam" diameter

$$\theta_p \cong \lambda_c / d_{tel} \approx 600 \text{ nm} / 25 \text{ mm} = 2.4 \times 10^{-5} \text{ rad} \approx 5 \text{ arcsec}. \qquad (2)$$

A spectrum is obtained for every spatial resolution element in the field. As noted in section II, the stroke length of the optical delay line sets the spectral resolution. If, for example, the physical stroke of the delay line is set to 50 μm, the optical delay $\Delta = 100$ μm and the spectral resolution,

$$R = \Delta / \lambda_c \approx 100 \text{ μm} / 600 \text{ nm} \approx 167. \qquad (3)$$

To understand how WIIT can image a field of view much wider than 5 arcseconds, next consider a source (or part of an extended scene) located at an angle θ relative to the optical axis, where $\theta \gg \theta_p$. Figure 6 illustrates schematically the relevant effects. If the angular offset is in a direction parallel to the baseline vector established by the collector mirrors, and we define this as the "x" direction, then light from the off-axis source will be brought to a focus on a pixel $(x_o + \delta x, y_o)$, where

$$\delta x = f_{lens} \theta(x_o + \delta x, y_o), \qquad (4)$$

and $f_{lens}$ is the focal length of the re-imaging lens. The lens is selected to focus the primary beam onto a 2 × 2 pixel area; in other words, the scene as viewed by a single sub-aperture mirror is critically sampled by the CCD camera. To be specific, $f_{lens} = 750$ mm so that when $\delta x = 18$ μm, the width of two pixels, $\theta \approx \theta_p \approx 5$ arcsec.

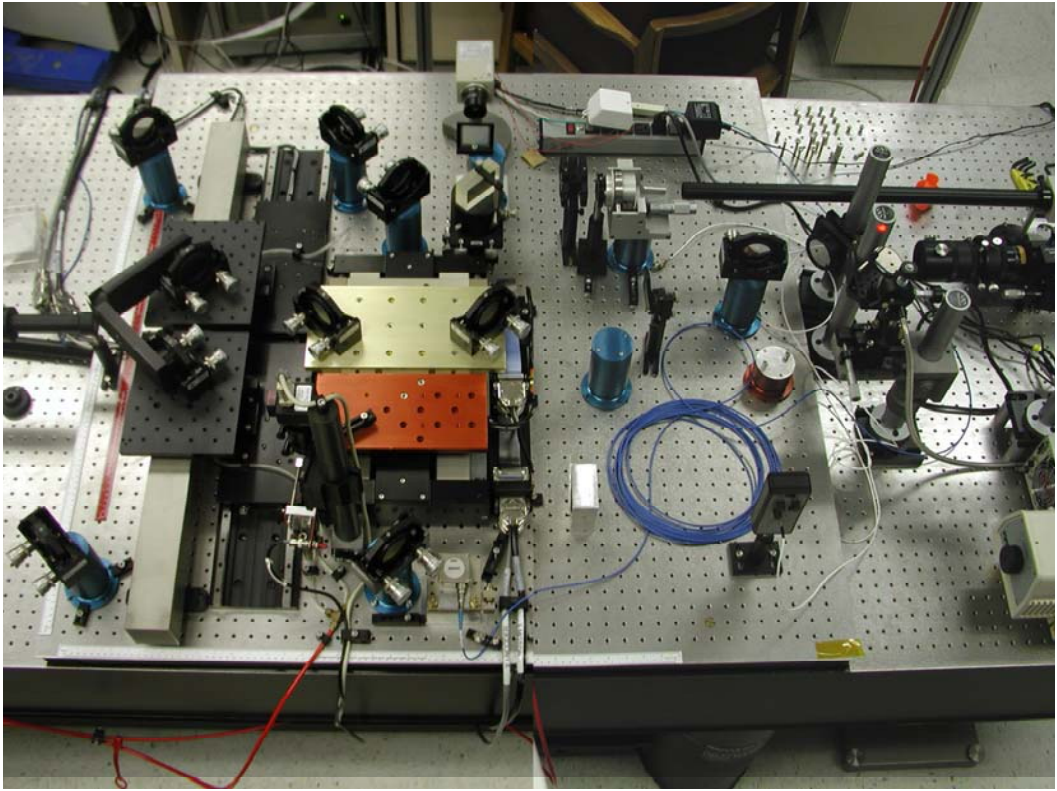

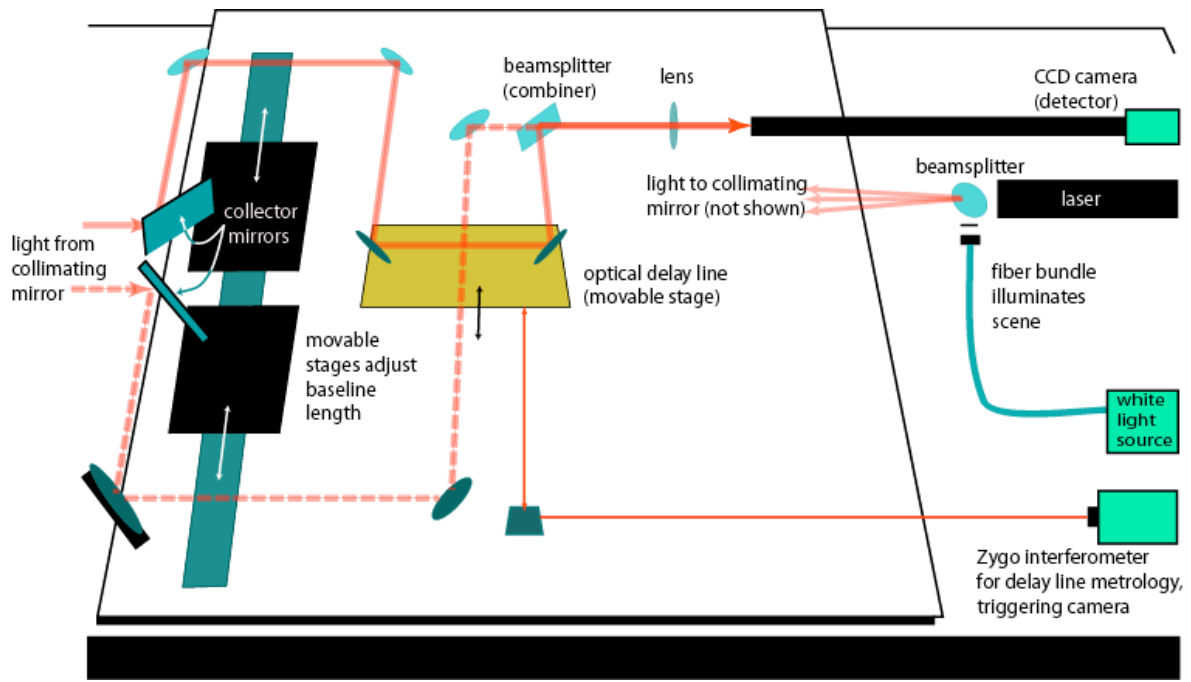

**Figure 3** (top) – Photograph of the Wide-field Imaging Interferometry Testbed.
**Figure 4** (bottom) – Layout of the testbed matching the photograph in Fig. 3.

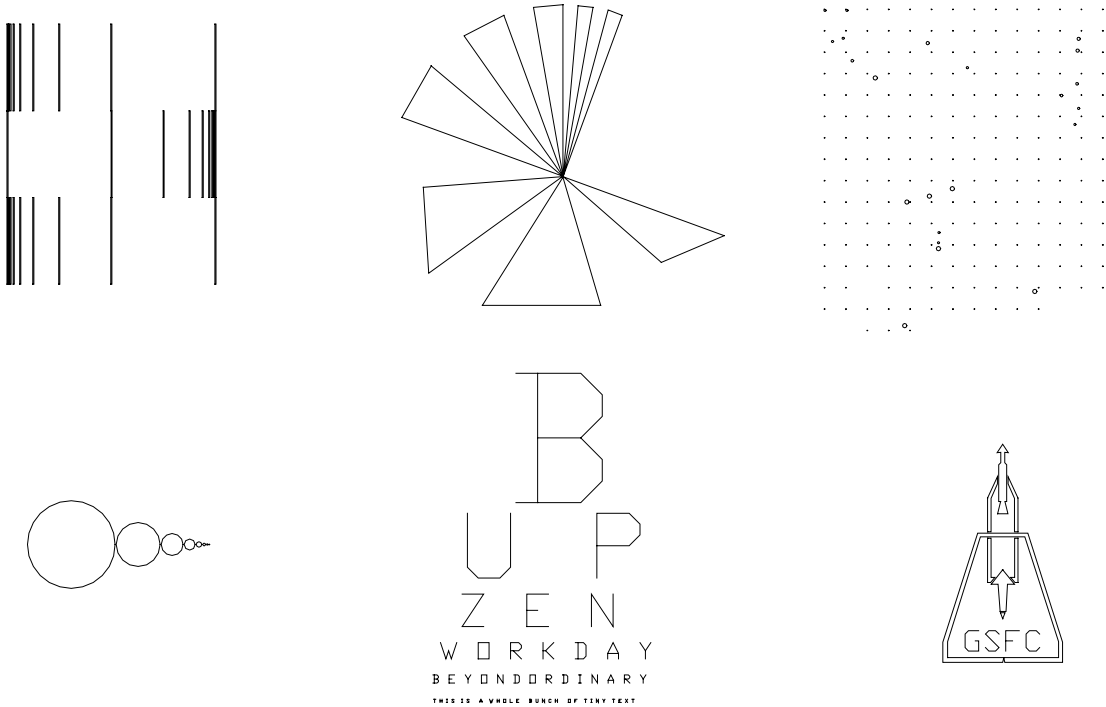

**Figure 5** – Test scenes ranging in complexity from simple pinholes to panchromatic images with complex spatial structure will be observed with the WIIT to measure resolution limits in both 1 and 2 dimensions, test mosaicing procedures, and measure the effect of cross-talk on synthesized image quality. This figure shows six representative images of the more than 100 scenes that we had microlithographically recorded on a glass plate. Each scene subtends several arcminutes and contains features on a variety of angular scales. The plate can be illuminated by a broadband white light source, and a bandpass filter can be used to simplify the spectral contents of the image. Ultimately, a realistic scene, such as the Hubble Deep Field (Fig. 1), can be used.

For an off-axis source located at field angle $\theta$ = 100 arcsec, $\delta x$ = 364 µm, which, given the pixel pitch, corresponds to a displacement by 40 pixels away from $(x_o, y_o)$. Inversely, the field of view covered by 100 × 100 pixels is a 4.1 arcmin square. The signal recorded as a function of time by a camera pixel located at $(x_o + \delta x, y_o)$ will be an interferogram whose Fourier transform is the spectrum of the off-axis source at the spatial frequency sampled by the baseline

$$b = b_o \cos\theta(x_o + \delta x, y_o) \approx b_o, \qquad (5)$$

where $b_o$ is the distance between the centers of the collecting mirrors (i.e., the baseline length for an on-axis source). In this case the ZPD will be displaced relative to its location for the on-axis source by an amount

$$\delta\Delta(x_o + \delta x, y_o) = b_o \sin\theta(x_o + \delta x, y_o). \qquad (6)$$

For $\theta$ = 100 arcsec and $b = b_{max}$ = 200 mm, $\delta\Delta$ = 97 µm. Note that this shift in the ZPD is comparable to the optical delay scan length required to provide modest spectral resolution. Therefore, by stroking the delay line through a distance somewhat greater than that required for spectroscopy, one can allow for the ZPD shift and obtain useful interferometric data for off-axis sources.

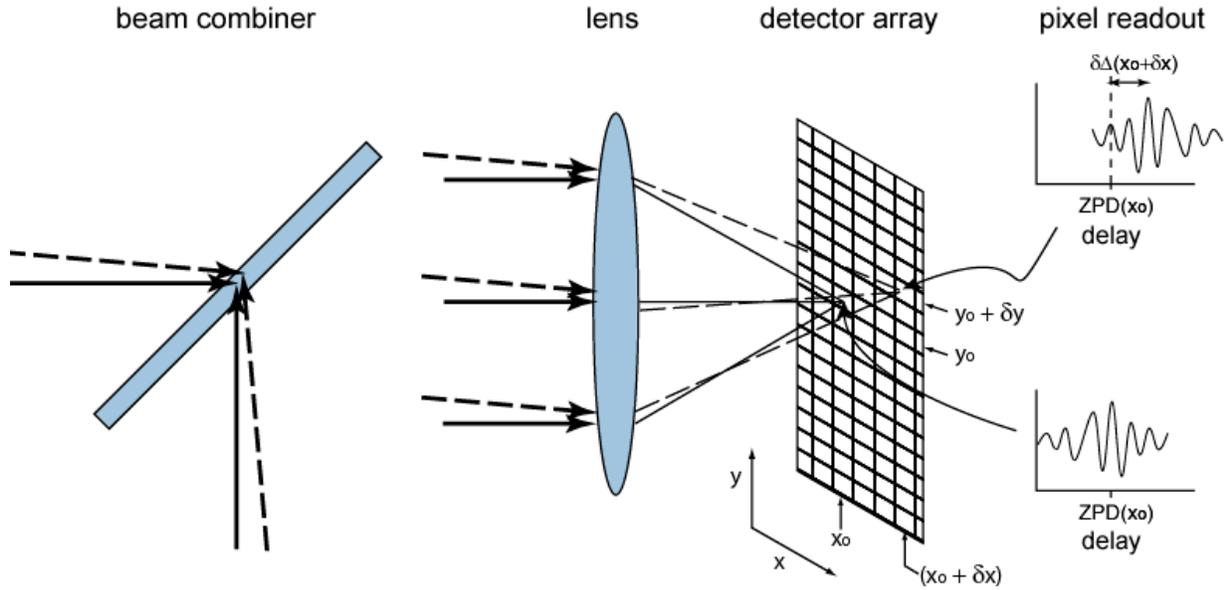

**Figure 6** – Light rays from a source located on the optical axis of the collimating mirror (solid lines), after passing along opposite arms of the interferometer, reach the beam combiner at right angles. A lens images those rays onto pixel $(x_o, y_o)$. As the optical delay line is scanned the pixel records an interferogram (bottom). Light rays from an off-axis source (dashed lines) strike the beam combiner symmetrically but at an acute or obtuse angle. These rays are imaged onto pixel $(x_o + \delta x, y_o + \delta y)$, which records an interferogram (top) whose ZPD is displaced by the delay $\delta\Delta$ relative to the ZPD for an on-axis source.

A relatively large ZPD shift was found in the case used for illustration above because we looked at a large field angle (100 arcsec), considered an offset along the x-axis (aligned with the baseline), and took $b = b_{max}$. For smaller field angles $\theta$, or shorter baselines, the ZPD shift $\delta\Delta$ would be smaller, and well within the optical delay range $\Delta = R\lambda_c$ covered to acquire spectroscopic data at resolution R for a source located on the optical axis. At field angles offset from the optical axis in the y direction (perpendicular to the baseline) there is no ZPD shift. In other words,

$$\delta\Delta(x_o, y_o + \delta y) = 0. \qquad (7)$$

Although the maximum baseline of SPECS (~1 km) is much greater than that of WIIT, the wavelength is proportionately longer and the same principle applies: by increasing the optical stroke length a modest amount, double Fourier data can be obtained for sources located at field angles $\theta \gg \theta_p$. If SPECS has 4 m diameter light collectors and is equipped with an $N_{pix} \times N_{pix}$ detector array, and if the array is used to Nyquist sample the primary beam, then the SPECS field of view at wavelength $\lambda$ would be

$$\theta_{FOV} \approx 4.3 \text{ arcmin } (N_{pix}/100) \, (\lambda/100 \text{ μm}) \, (d_{tel}/4 \text{ m})^{-1}, \qquad (8)$$

which is about twice the diameter of the Hubble Deep Field.

Above we gave the theory that underlies the performance of an ideal incarnation of WIIT. Environmental conditions, imperfect components, imperfect alignment, motion smearing, and uncertainties associated with various measurements (e.g., metrology, camera read noise, etc.) affect the actual performance of the testbed. Discussion of these effects is beyond the scope of this paper, but can be found in Papers II and III. Here we simply note that an important element of our experimental approach is to measure and model the performance of the interferometer and its various components, and to measure and minimize environmental effects, such as air turbulence, in an effort to understand the various error sources and their relative impacts on overall system performance. For example, we have found that wavefront distortion associated with air turbulence is one of the most severe effects, but that even rudimentary enclosures yield significant improvements in data quality (Paper II). If these improvements should turn out to be inadequate the entire testbed can be moved into the stable environment of the DGEF chamber; indeed, the location of the WIIT lab was chosen with this possibility in mind. By taking this approach we are able to make informed decisions on how to invest limited resources to improve WIIT, and we can gain valuable insight that can be applied to the design of interferometers for space.

### 4.2.  The Data

Raw data from the WIIT consist of a series of CCD camera frames corresponding to the sampled optical path lengths and baselines, and ancillary data. The ancillary data include real-time measurements of the instrument and its environment, as well as asynchronous measurements, such as mirror, filter, and beamsplitter spectral response functions, direct digital images of the test scenes, collector mirror rail straightness and alignment data, focus data, data gathered for independent verification of the performance of the optical encoders, measurements of light intensity as a function of position along the collector mirror rail, and as a function of time to verify long-term source brightness stability, long-period temperature trend data, and measurements of the effect of air turbulence on optical path length along various paths across the testbed optical table. In concert with the science data (camera frames), readings are taken from sixteen temperature sensors placed at strategic locations throughout the testbed, and from a detector used to monitor source brightness incident on the interferometer at about 1 Hz. Absolute metrology data from optical encoders located at the two collector mirrors, on the optical delay stage, and on the rotation stages used to rotate the test scene and the camera are also captured in real time.

To facilitate data reduction and analysis the science data and synchronously collected ancillary data are recorded in the astronomical standard FITS (Flexible Image Transport System) format. All of the data from a single delay line scan are recorded in a single FITS file. The header and housekeeping data records in the file include all the information needed to calculate the baseline vector.

A full science data set is voluminous, even when only a subset of the CCD camera data are saved. The field of view whose angular dimensions are $\theta_x \times \theta_y$ is covered by $N_x \times N_y$ pixels, where

$$N_x \approx 2\theta_x \, d_{tel} / \lambda_c \text{ and} \qquad (9a)$$
$$N_y \approx 2\theta_y \, d_{tel} / \lambda_c, \qquad (9b)$$

and the angles are given in radians. If the total optical delay scan length is set to allow for a desired spectral resolution $R_d$ and the ZPD shift associated with a source at an off-axis field angle $\theta_x/2$, and if $f_s$ samples (camera frames) are obtained per fringe, then the number of frames per delay line scan

$$N_f = f_s (R_d + \delta\Delta / \lambda_c), \qquad (10)$$

where $\delta\Delta \approx b_{max}\theta_x/2$ is the maximum possible ZPD shift. To estimate the number of baseline measurements required for high-quality image synthesis we assume that the synthetic aperture will be uniformly covered and critically sampled by the $d_{tel}$ diameter collector mirrors. Thus, the number of baseline measurements, and therefore the number of FITS files, is given conservatively by

$$N_b = 4(b_{max} / d_{tel})^2. \qquad (11)$$

The number of bytes of data in a complete data set is therefore given by

$$B = b_{pix} \, N_x \, N_y \, N_f \, N_b, \qquad (12)$$

where $b_{pix}$ is the number of bytes per pixel.

When WIIT ($d_{tel}$ = 25 mm; $\lambda_c$ = 0.6 µm; $f_s$ = 4; $b_{pix}$ = 2) images a 4 arcmin square field of view at spectral resolution $R_d$ = 1000, and with $b_{max}$ set to provide 0.3 arcsec angular resolution, 25 GB of data are produced. After removal of the instrument signature these data can be Fourier transformed to yield a synthesized "data cube," which has two spatial dimensions and one spectral dimension. This data cube can be compared with a high-resolution direct digital image of the test scene used to generate the interferometric data to evaluate the performance of the testbed and the image synthesis algorithms.

### 4.3. The Algorithms

We intend to explore a variety of possible data reduction algorithms. In so doing we shall make extensive use of the principles and techniques developed over five decades by radioastronomers.[2]

Conceptually, the simplest algorithm views the data as if they were produced by $N_x \times N_y$ separate Michelson interferometers. The method works one pixel at a time. Here we use the word "pixel" somewhat loosely. The WIIT camera can be rotated to keep the image stationary on the focal plane when the test scene is rotated, in which case any given off-axis source will always be imaged onto the same camera pixel. However, if the test scene is *not* de-rotated at the focal plane, then the pixel to which we are referring is the camera pixel that corresponds to a fixed off-axis angle in the test scene, which generally changes as a function of time. The first step involves Fourier transforming the 1-dimensional $N_f$-channel interferogram measured for each baseline to produce a spectrum, and requires a total of $N_x \times N_y \times N_b$ Fourier transforms. The resulting data are reorganized into $N_x \times N_y$ "u-v data sets," each in a separate file, where each file contains the data from a single pixel, but for all baselines. Next the data in each file are reduced one spectral channel at a time. For each spectral channel a 2-D Fourier transform yields an image of a small part of the test scene, specifically a region subtended by a single primary beam. The image can be improved by applying the CLEAN algorithm[19] to clean up the "dirty" synthesized beam. At the end of this process there are $N_x \times N_y$ data cubes, each corresponding to a different region in the wide-field test scene, with overlapping spatial coverage. By stitching together the individual data cubes, a single data cube, which represents the spatial and spectral content of the wide-field test scene, can be created. The synthetic beam gives the spatial resolution, and the spectral resolution corresponds to the range of optical delay sampled in each beam. This algorithm does not fully exploit the information available in the raw data because correlations due to the critically sampled primary beam and the oversampled u-v plane are not properly taken into account.

More sophisticated algorithms will allow for the correlations between data samples and known properties of the source (e.g., positive brightness) and the instrument (e.g., baseline and delay line metrology). Each individual data sample – the calibrated intensity measured in a single camera pixel in a single CCD frame – represents a particular location in the test scene (sampled at the diffraction limited resolution of the $d_{tel}$ = 25 mm WIIT mirrors), at a particular optical delay relative to ZPD, and at a particular spot in the u-v plane determined by the baseline length and orientation, and there are measurement errors or uncertainties associated with all of these quantities.

A second possible algorithm would iteratively fit a model brightness distribution to minimize the difference between the expected response of the interferometer to this brightness distribution and the observed response, subject to the constraints, and given the measurement uncertainties and correlations (i.e., data covariance matrix). The expected response of an ideal interferometer to a quasi-monochromatic incoherent source over a small field is the two-dimensional Fourier transform of the spatial brightness distribution, and this "van Cittert-Zernike theorem" can be generalized for a non-ideal interferometer, a polychromatic source, and a field of finite extent.[20] The brightness distribution could be fitted either one spectral channel at a time or simultaneously at all wavelengths in the spectral bandpass. The latter approach might be better, as it would allow for the introduction of additional physical constraints (e.g., well-understood emission mechanisms, such as thermal emission from interstellar dust and emission in spectral lines of known rest wavelengths in the far-infrared and submillimeter). Needless to say, this would be theoretically and computationally ambitious, but it is possible. Model fitting also has a heritage in radioastronomy.[21]

A third approach, better than the first one described above and much easier to implement than the second, would involve interpolation of the data onto an even u-v and wavenumber grid prior to discrete Fourier inversion into a spatial-spectral data cube. Direct Fourier inversion, without special gridding, is also possible, and the visibility data could be weighted to improve the shape of the synthesized beam. The results of this method could serve as the starting point for iteration if the second, iterative approach is developed. Alternatively, CLEAN[19] or the Maximum Entropy Method [22] could be used to further improve image quality.

## 5. SUMMARY

Far-IR and submillimeter imaging interferometry has the potential to revolutionize our understanding of galaxy, star, and planet formation. Recent advances in far-IR detector technology are such that background-limited performance with cryogenic telescopes is within reach, but detector arrays will be limited for the foreseeable future to $\sim 10^4$ pixels, making image plane beam combination impractical. Michelson (pupil plane) beam combination is the natural solution in this case. A Michelson interferometer operated in "double Fourier" mode can yield valuable spectroscopic information as well as high-resolution imagery. The WIIT project is designed to overcome a limitation often ascribed to Michelson interferometers: their limited field of view. In principle, the pixels in a detector array can be used to record simultaneously interferometric data from many contiguous locations in the sky. The Wide-field Imaging Interferometry Testbed will be used to develop the technique of spatially multiplexed double Fourier optical/IR interferometry. One pays a penalty in the time domain for operating an interferometer in this manner, but the penalty is small when the optical delay line is already being scanned to obtain spectral data. The data from WIIT will be used to test algorithms for wide-field mosaic imaging. If this technique can be perfected it will enable future far-IR and submillimeter interferometers to image fields several arcminutes in diameter at high resolution, and provide a spectrum at each field location.

## ACKNOWLEDGMENTS

Funding for WIIT is provided by NASA Headquarters through the ROSS/SARA Program and by the Goddard Space Flight Center through the IR&D Program. The SPIRIT and SPECS studies receive funding support from NASA Headquarters through the RASC program. We appreciate the expert advice provided by members of the WIIT Science and Technical Advisory Group, whose members are Richard Burg, Bill Danchi, Dan Gezari, Antoine Labeyrie, John Mather (Chair), Harvey Moseley, Dave Mozurkewich, Peter Nisenson, Stan Ollendorf, Mike Shao, and Hal Yorke. Dr. Mundy was a National Research Council Senior Research Associate at NASA GSFC from October 2001 to June 2002.